\documentclass[fleqn,12pt]{wlscirep}
\usepackage[utf8]{inputenc}
\usepackage[T1]{fontenc}
\usepackage{xcolor}
\usepackage[font=scriptsize]{caption}
\usepackage[version=4]{mhchem}
\usepackage{pifont}
\usepackage{chemformula}
\usepackage{filecontents}
\usepackage[left]{lineno}

\usepackage[normalem]{ulem}

\newcommand{\change}[1]{#1}           
\newcommand{\aj}{Astron. J.}   
\newcommand{\apj}{Astrophys. J.}   
\newcommand{\apjl}{Astrophys. J. Lett.}   
\newcommand{\aap}{Astron. Astrophys.}   
\newcommand{\icarus}{Icarus}   
\newcommand{\joss}{J. Open Source Softw.} 
\newcommand{\mnras}{Mon. Not. R. Astron. Soc.}   
\newcommand{\nastro}{Nat. Astron.} 
\newcommand{\pasp}{Publ. Astron. Soc. Pac.}   
\newcommand{\sci}{Science} 
\title{Water in the terrestrial planet-forming zone of the PDS~70 disk}

\author[1,*]{G. Perotti}
\author[2]{V. Christiaens}
\author[1]{Th. Henning}
\author[3]{B. Tabone}
\author[4,5]{L. B. F. M. Waters}
\author[6]{I. Kamp}
\author[7]{G. Olofsson}
\author[8]{S. L. Grant}
\author[9]{D. Gasman}
\author[1]{J. Bouwman}
\author[1]{M. Samland}
\author[1]{R. Franceschi}
\author[8,10]{E. F. van Dishoeck}
\author[1]{K. Schwarz}

\author[1,11,12]{M. G\"udel}
\author[13]{P.-O. Lagage}
\author[14]{T. P. Ray}
\author[9]{B. Vandenbussche}

\author[3]{A. Abergel}
\author[2]{O. Absil}
\author[6]{A. M. Arabhavi}
\author[9]{I. Argyriou}
\author[15]{D. Barrado}
\author[16]{A. Boccaletti}
\author[14,17]{A. Caratti o Garatti}
\author[18]{V. Geers}
\author[12]{A. M. Glauser}
\author[19]{K. Justannont}

\author[20]{F. Lahuis}
\author[6]{M. Mueller}
\author[13]{C. Nehm\'e}
\author[13]{E. Pantin}
\author[1]{S. Scheithauer}
\author[8]{C. Waelkens}

\author[11]{R. Guadarrama}
\author[4]{H. Jang}
\author[6,21,22]{J. Kanwar}
\author[15]{M. Morales-Calder\'on}
\author[11]{N. Pawellek}

\author[14]{D. Rodgers-Lee}
\author[1]{J. Schreiber}

\author[23]{L. Colina}
\author[24]{T. R. Greve}
\author[25]{G. \"Ostlin}
\author[18]{G. Wright}
\affil[1]{Max Planck Institute for Astronomy, K{\"o}nigstuhl 17, D-69117 Heidelberg, Germany}

\affil[2]{STAR Institute, Universit\'e de Li\`ege, All\'ee du Six Ao\^ut 19c, 4000 Li\`ege, Belgium}

\affil[3]{Universit\'e Paris-Saclay, CNRS, Institut d’Astrophysique Spatiale, 91405, Orsay, France}

\affil[4]{Department of Astrophysics/IMAPP, Radboud University, PO Box 9010, 6500 GL Nijmegen, The Netherlands}

\affil[5]{SRON Netherlands Institute for Space Research, Niels Bohrweg 4, NL-2333 CA Leiden, the Netherlands}

\affil[6]{Kapteyn Astronomical Institute, Rijksuniversiteit Groningen, Postbus 800, 9700AV Groningen, The Netherlands}

\affil[7]{Department of Astronomy, Stockholm University, AlbaNova University Center, 10691 Stockholm, Sweden}

\affil[8]{Max-Planck Institut f\"{u}r Extraterrestrische Physik (MPE), Giessenbachstr. 1, 85748, Garching, Germany}

\affil[9]{Institute of Astronomy, KU Leuven, Celestijnenlaan 200D, 3001 Leuven, Belgium}

\affil[10]{Leiden Observatory, Leiden University, 2300 RA Leiden, the Netherlands}

\affil[11]{Dept. of Astrophysics, University of Vienna, T\"urkenschanzstr 17, A-1180 Vienna, Austria}

\affil[12]{ETH Z\"urich, Institute for Particle Physics and Astrophysics, Wolfgang-Pauli-Str. 27, 8093 Z\"urich, Switzerland}

\affil[13]{Universit\'e Paris-Saclay, Universit\'e Paris Cit\'e, CEA, CNRS, AIM, F-91191 Gif-sur-Yvette, France}

\affil[14]{Dublin Institute for Advanced Studies, 31 Fitzwilliam Place, D02 XF86 Dublin, Ireland}

\affil[15]{Centro de Astrobiolog\'ia (CAB), CSIC-INTA, ESAC Campus, Camino Bajo del Castillo s/n, 28692 Villanueva de la Ca\~nada,
Madrid, Spain}

\affil[16]{LESIA, Observatoire de Paris, Universit\'e PSL, CNRS, Sorbonne Universit\'e, Universit\'e de Paris, 5 place Jules Janssen, 92195 Meudon, France}

\affil[17]{INAF – Osservatorio Astronomico di Capodimonte, Salita Moiariello 16, 80131 Napoli, Italy}

\affil[18]{UK Astronomy Technology Centre, Royal Observatory Edinburgh, Blackford Hill, Edinburgh EH9 3HJ, UK}

\affil[19]{Chalmers University of Technology, Onsala Space Observatory, 439 92 Onsala, Sweden}

\affil[20]{SRON Netherlands Institute for Space Research, PO Box 800, 9700 AV, Groningen, The Netherlands}

\affil[21]{Space Research Institute, Austrian Academy of Sciences, Schmiedlstr. 6, A-8042, Graz, Austria}

\affil[22]{TU Graz, Fakultät für Mathematik, Physik und Geodäsie, Petersgasse 16 8010 Graz, Austria}

\affil[23]{Centro de Astrobiolog\'ia (CAB, CSIC-INTA), Carretera de Ajalvir, E-28850 Torrej\'on de Ardoz, Madrid, Spain}

\affil[24]{DTU Space, Technical University of Denmark, Building 328, Elektrovej, 2800 Kgs. Lyngby, Denmark}

\affil[25]{Department of Astronomy, Oskar Klein Centre; Stockholm University; SE-106 91 Stockholm, Sweden}


\begin{document}
\flushbottom
\maketitle
\vspace{-10pt}
\textbf{Terrestrial and sub-Neptune planets are expected to form in the inner ($<10~$AU) regions of protoplanetary disks.\cite{Mulders2018} Water plays a key role in their formation,\cite{Ciesla2006,Eistrup2022,Krijt2022} although it is yet unclear whether water molecules are formed in-situ or transported from the outer disk.\cite{Bethell2009,Glassgold2009} So far \textit{Spitzer Space Telescope} observations have only provided water luminosity upper limits for dust-depleted inner disks,\cite{Banzatti2020} similar to PDS~70, the first system with direct confirmation of protoplanet presence.\cite{Keppler2018,Haffert2019} Here we report JWST observations of PDS~70, a benchmark target to search for water in a disk hosting a large ($\sim54~$AU) planet-carved gap separating an inner and outer disk.\cite{Long2018,Keppler2019} Our findings show water in the inner disk of PDS~70. This implies that potential terrestrial planets forming therein have access to a water reservoir. The column densities of water vapour suggest in-situ formation via a reaction sequence involving O, H$_2$, and/or OH, and survival through water self-shielding.\cite{Bethell2009} This is also supported by the presence of CO$_2$ emission, another molecule sensitive to UV photodissociation. Dust shielding, and replenishment of both gas and small dust from the outer disk, may also play a role in sustaining the water reservoir.\cite{Benisty2021} Our observations also reveal a strong variability of the mid-infrared spectral energy distribution, pointing to a change of inner disk geometry.}

\begin{figure}[htb!]
\centering
\includegraphics[width=\hsize]{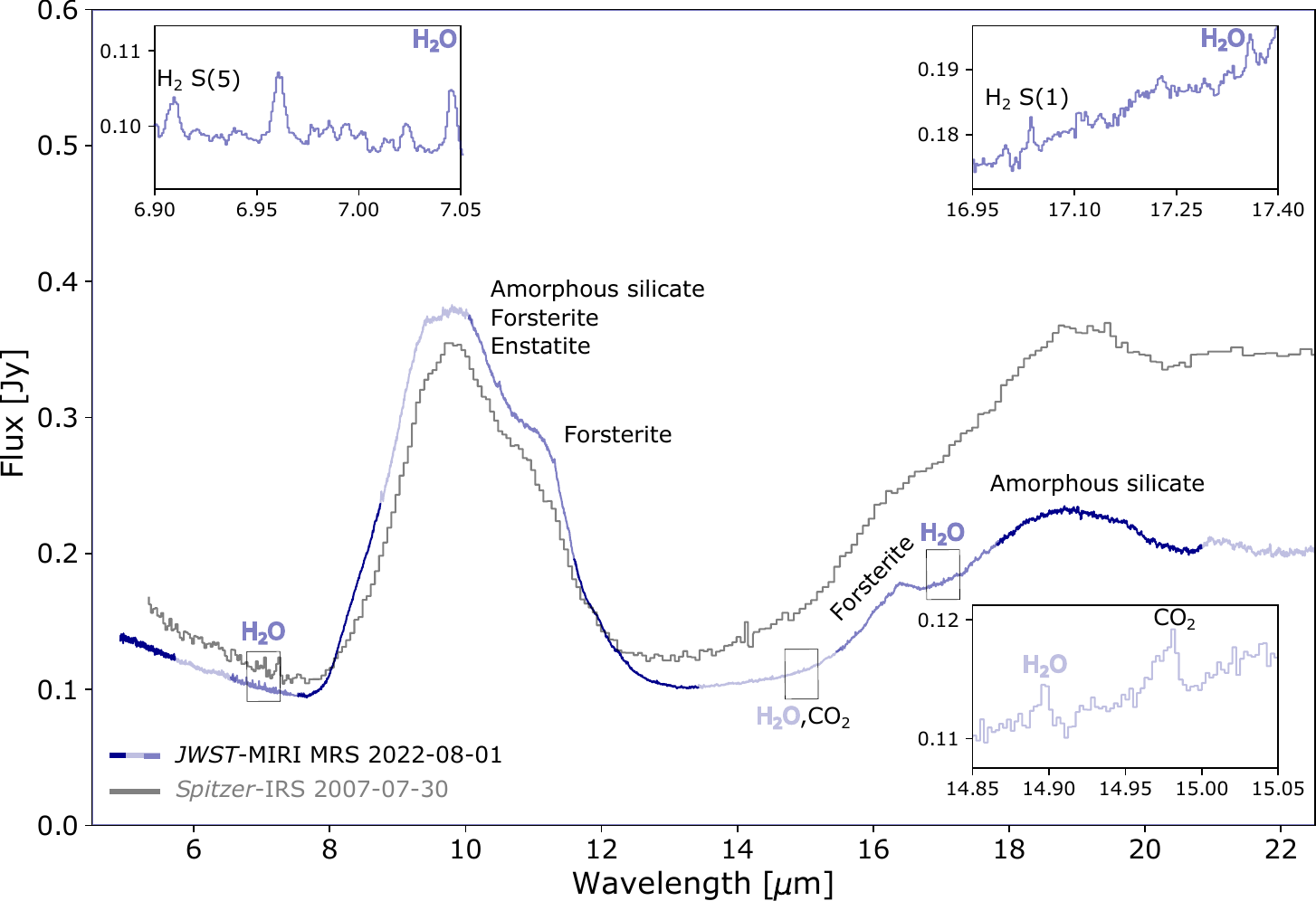} 
\caption{\textbf{JWST-MIRI-MRS spectrum of PDS~70.} The spectrum is a composite of the colour-coded short, medium, and long sub-bands of the four MIRI-MRS Integral Field Units (IFUs).\cite{miri_rieke2015PASP} The \textit{Spitzer}-IRS spectrum of PDS~70 is also shown in grey. \change{The major dust features are labelled. The spectrum is dominated by exceptionally prominent silicate emission at 10 and 18~$\mu$m and it clearly shows a number of crystalline dust features. The much higher sensitivity and spectral resolution of MIRI-MRS compared with \textit{Spitzer-IRS} allows us to detect for the first time an inner disk gas reservoir by showing weak emission of water vapour and carbon dioxide as well as two molecular hydrogen lines.} The insets show the ro-vibrational and rotational transitions of ortho- and para-\ce{H2O}, the molecular hydrogen H$_2$ $S$(1) and $S$(5) rotational lines, and the $\nu_5$ bending mode of \ce{CO2}.} 
\label{fig1}
\end{figure}

\thispagestyle{empty}
\smallskip 
Observations of PDS 70 were taken with the JWST Mid-InfraRed Instrument (MIRI)\cite{miri_rieke2015PASP,Wright2015} Medium Resolution Spectrometer\cite{Wells2015} (MRS; spectral resolving power $R \sim~1600-3400$) as part of the guaranteed time MIRI mid-INfrared Disk Survey (MINDS; \change{see Methods and Extended Data Fig.~\ref{figA1}).} The complete spectrum of PDS~70 shows several distinct traits (Fig.~\ref{fig1}), which stands out with respect to other T~Tauri disks.\cite{Kessler-Silacci2006,Furlan2006}

A significant flux offset - up to a factor of 1.5 at wavelengths $\geq18~\mu$m - is found between the \change{MIRI and the archival \textit{Spitzer} InfraRed Spectrograph (IRS) low-resolution ($R \sim~$$60-100$) spectra} recorded with 15 years and one day time difference. This discrepancy is too large to be explained by calibration uncertainties; the absolute uncertainty for both IRS and MIRI is $\sim5\%$ for the $4.9-22.5\,\mu$m range.\cite{Argyriou2023} Similarly, the difference in aperture size of the two \change{spectrographs} cannot account for such an offset. Hence, with the current MIRI data reduction, time variability is the most likely explanation for the observed flux differences.

Variability in the mid-IR observed with \textit{Spitzer}-IRS has been mainly attributed to short-wavelength stellar irradiation or to dynamical changes in the inner disk geometry due to the presence of planets.\cite{Muzzerolle2009,Espaillat2011} 
In the case of PDS~70, stellar irradiation is excluded as it would cause an overall increase or decrease in flux contrary to what is observed with the Wide-field Infrared Survey Explorer (WISE) time-series observations (Extended Data Fig.~\ref{figA6}). PDS~70 is known to be in a late stage of accretion - with an estimated waning accretion rate\cite{Manara2019,Skinner2022} of $\sim 10^{-10}\,M_\odot$~yr$^{-1}$ making it unlikely to explain the significant flux difference. Changes in the scale height of the inner disk wall emitting at shorter wavelengths ($\sim2-8~\mu$m) can be responsible for shadowing the disk material located further out, resulting in less emission at wavelengths beyond $18-20~\mu$m (also referred to as "seesaw-like" variability\cite{Muzzerolle2009}). However, for PDS~70 a complete seesaw-like profile is not observed as there is no corresponding increase in flux at the shorter wavelengths. Time variability is also supported by the aforementioned WISE observations which indicate that such variability occurs on short timescales ($\leq1~$yr) and that it may indeed be attributed to occulting material located close to the star ($\sim1~$AU). 

\begin{figure}[htb!]
\centering
\includegraphics[width=14cm]{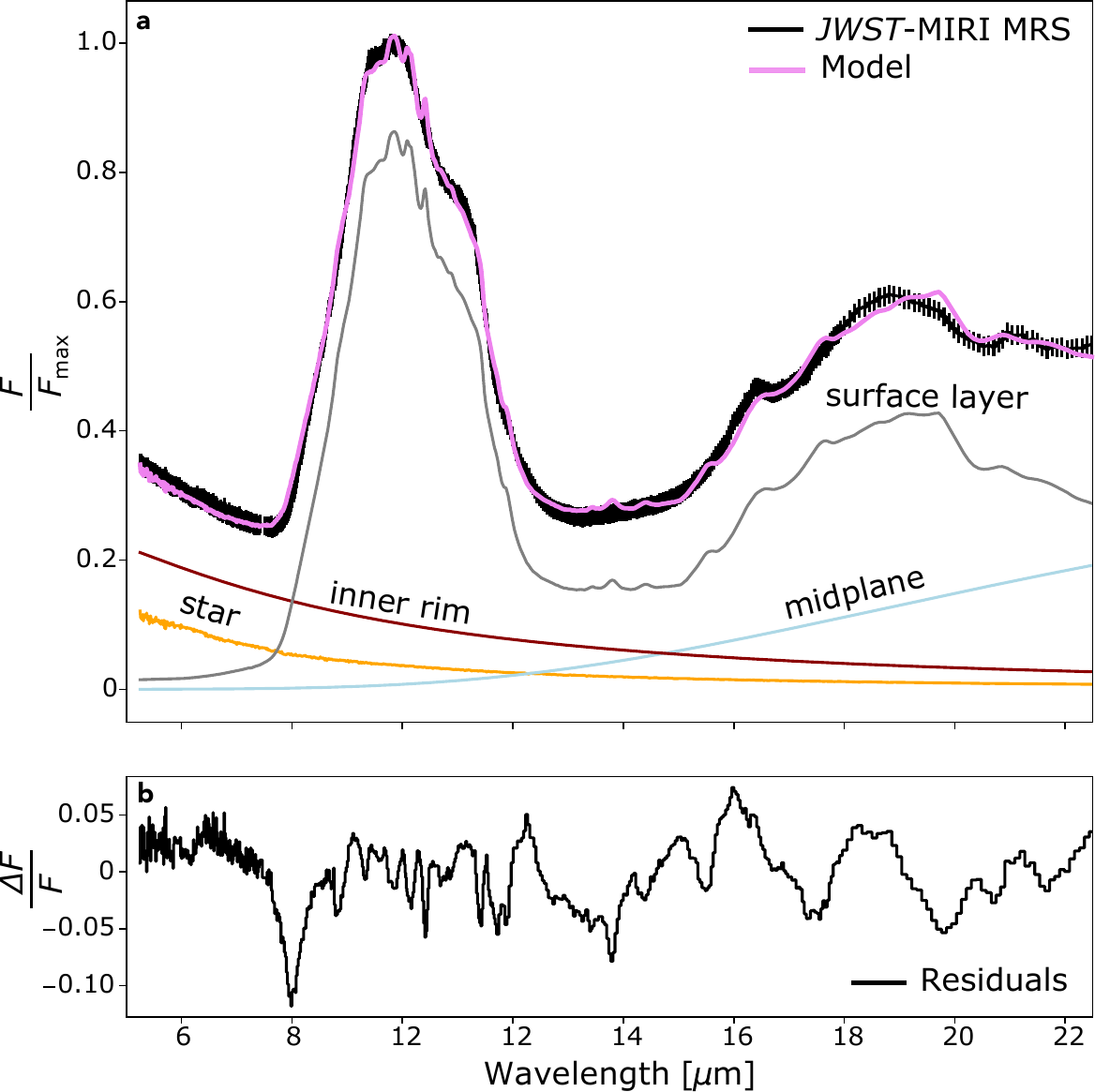}
\caption{\textbf{Dust continuum fit to the MIRI spectrum of PDS~70.} \textbf{a,} The disk model has three spectral components: an inner rim, an optically thick mid-plane disk layer, and an optically thin warm disk surface layer. The stellar photospheric emission is represented by a stellar atmosphere model \change{(see Methods for further details)}. The surface layer component dominates the MIRI spectrum in the $8-22.5\,\mu$m wavelength range. Its temperature is constrained to be between 400 and 600~K. The silicate emission at $8-12~\mu$m is consistent with a population of optically thin dust grains with typical sizes of $0.1-2~\mu$m. A significant contribution from an optically thick dust component is excluded because of the high silicate peak/continuum ratio of $\simeq~$4.\cite{Oliveira2010} \textbf{b,} The residuals on the dust continuum fit.}
\label{fig2}
\end{figure}

The MIRI spectrum of PDS~70 clearly shows the presence of silicate dust grains that have undergone significant thermal processing (Fig.~\ref{fig1}). We attribute the crystalline dust features to enstatite  at 9.40$~\mu$m and forsterite at 11.30$~\mu$m and 16.40$~\mu$m. The observed dust continuum is well reproduced with a three-component disk model, with a $400-600~$K surface layer accounting for the bulk of the observed emission (Fig.~\ref{fig2}). 

\begin{figure}[htb!]
\centering
\includegraphics[width=\hsize]{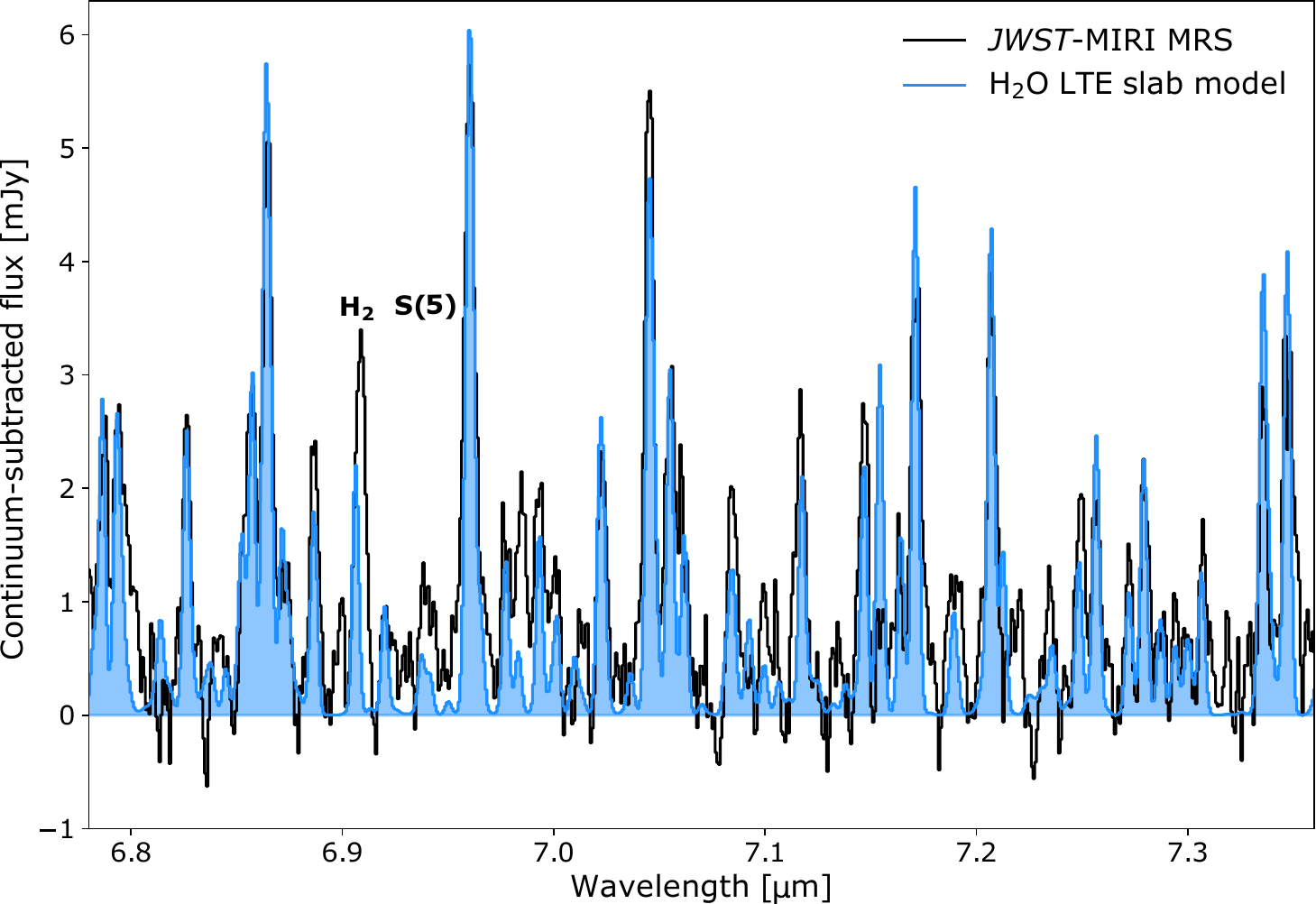}
\caption{\textbf{Continuum-subtracted spectrum showing H$_2$O emission in the 7~$\mu$m region and the best-fit LTE slab model.} The best-fit model (blue) has $T=600~$K, $N\mathrm{(H_2O)} = 1.4~\times10^{18}~$cm$^{-2}$, and $R=0.047~$AU. The molecular hydrogen H$_2$ $S$(5) line is labelled on top of the spectrum.}
\label{fig3}
\end{figure}

The MIRI spectrum also reveals a wealth of water lines, particularly in the 7~$\mu$m spectral window (Fig.~\ref{fig3}). 
This indicates the presence of a water reservoir in the terrestrial region of a disk already hosting two or more protoplanets. As such, it also provides important clues to theories on the origin of water during terrestrial planet formation in the solar system.\cite{Stimpfl2006,Genda2008} We focus on the ro-vibrational transitions of the bending mode of para- and ortho-water in the 7~$\mu$m region where the brightest lines are observed and contamination by the stellar atmosphere is negligible (Extended Data Fig.~\ref{figA3}). This includes strong water blends dominated by lines with upper energy levels $E_u \simeq 2400-3200~$K. Weaker lines are also detected at the 1~mJy level, some of them corresponding to more excited levels up to $E_u \simeq 4300~$K.

Further insight into the origin of water emission is obtained from zero-dimensional slab modeling which also has been used to interpret \textit{Spitzer} spectroscopic data.\cite{Salyk2011} The synthetic spectrum of water is calculated from a plane-parallel slab model, where the level populations are in Local Thermodynamic Equilibrium (LTE) at a single excitation temperature $T$. The other fitting parameters are the line-of-sight column density $N$ within an effective emitting area $\pi R^2$ given by its radius $R$, and the intrinsic line broadening assumed to be $\sigma = 2~$km/s.\cite{Salyk2011} Note that $R$ does not need to correspond to a disk radius, but could also represent an annulus with the same area or an emission spot breaking the axisymmetry. The best-fit model is then obtained by minimizing the reduced $\chi^2$ between measured and model line fluxes over the individual spectral window around each H$_2$O line (Extended Data Fig.~\ref{figA4}). 

The observed H$_2$O spectrum in the 6.78$-$7.36~$\mu$m spectral region is best fitted with a slab of gas at $T=600~$K, with an emitting area of radius $R=0.047~$AU, and a column density of $N=1.4 \times 10^{18}$cm$^{-2}$.
The temperature is mostly determined by the ratio between lines of different $E_u$, for example, the series of lines in the 7.3$~\mu$m region. The column density is set by the ratio between the weaker lines and indicates that the brightest lines are optically thick. The emitting area is constrained by matching the fluxes of the optically thick lines, and points toward a compact emission region. This is further supported by the fact that the detected lines are broad ($\Delta \lambda \sim 0.01-0.05~\mu$m); if the line broadening is caused by the gas kinematics, the FWHM of the line would be about $100~$km/s, corresponding to a Keplerian radius of 0.1~AU, consistent with the emitting area deduced from our fit after correction for disk inclination $i=51.7 \pm 0.1^\circ$.\cite{Keppler2019} 
Interestingly, we find that our best-fit LTE model of the water emission in the 7$~\mu$m region reproduces reasonably well water rotational lines at 15$~\mu$m, 
suggesting that all water emission in the MIRI spectral range originates from inside $\sim0.05$~AU under LTE conditions (Extended Data Fig.~\ref{figA5}).  

Besides water in the 7$~\mu$m, 15$~\mu$m, and 17$~\mu$m regions, other species have been identified whose analysis is postponed to a future study. \textbf{CO$_2$:} The fundamental Q-branch of CO$_2$ corresponding to the $\nu_5$ bending mode is detected at 14.96~$\mu$m (Extended Data Fig.~\ref{figA5}). Interestingly, the width of this feature is sensitive to the temperature and indicates cooler gas of $T \simeq 200~$K in the optically thin regime. \textbf{H$_2$:} The pure rotational molecular hydrogen H$_2$ $S(5)$ and H$_2$ $S(1)$ lines are detected at 6.91$~\mu$m and 17~$\mu$m (Fig.~\ref{fig1}). We note that the H$_2$ $S(2)$, $S(3)$, and $S(4)$ lines coincide with the broad silicate emission feature and thus establishing their presence needs to await an in-depth analysis of this dust feature.  

\begin{figure}[ht!]
\centering
\includegraphics[width=\hsize]{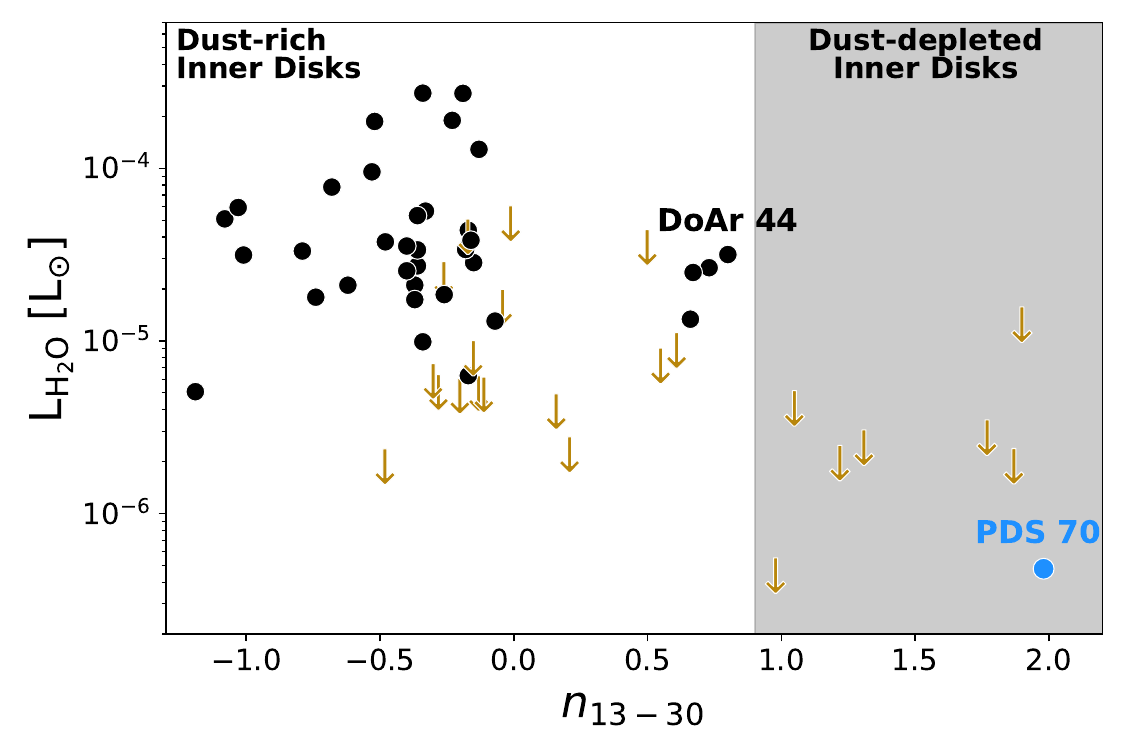}
\caption{\textbf{Comparison between water luminosity and mid-infrared spectral index ($n_\mathrm{13-30}$) for a sample\cite{Banzatti2020} of protoplanetary disks.}
\change{$n_\mathrm{13-30}$ is a diagnostic of the presence and size of inner disk dust cavities: $0.9 < n_\mathrm{13-30} < 2.2$ corresponds to disks with large gaps and/or cavities.\cite{Brown2007,Furlan2009}} Black dots represent disks with mid-infrared water detections. Disks for which only upper limits were obtained are shown as \change{gold} arrows. The grey shaded area highlights the location of dust-depleted inner disks with PDS~70 shown as a blue dot. The water line flux used to compute the water luminosity of PDS~70 is calculated as described in a previous work.\cite{Banzatti2020} The \textit{Spitzer} spectrum is used to estimate $n_\mathrm{13-30}$ for PDS~70 to be consistent with the other targets. \change{\textit{Spitzer}-IRS obtained only water luminosity upper limits for disks characterized by $n_\mathrm{13-30}$ greater than 0.9.} Below 10~$\mu$m, IRS provided only a spectral resolution of $R\,\sim100$, preventing a comprehensive view of water in the innermost regions.\cite{Banzatti2023} DoAr~44 is a system schematically similar to PDS~70.\cite{Salyk2015} The two stars have comparable age and spectral types K3 and K7 respectively; \change{DoAr~44 has a higher mass accretion rate of $\dot{M}_\mathrm{acc} \sim 10^{-8} \,M_\odot$~yr$^{-1}$.\cite{Manara2014} The cavity size of DoAr~44 is 34~AU\cite{Salyk2015,Banzatti2020}, smaller than that of PDS~70 ($\sim$54~AU\cite{Keppler2019}). 
Both systems have small inner disks based on VLTI-GRAVITY\cite{Bouvier2020}, VLT-SPHERE and ALMA data \cite{Keppler2018,Keppler2019}, and in both systems, the water emission is contained to within 1~AU.\cite{Salyk2015} The water luminosity of PDS~70 is two orders of magnitude weaker than that of DoAr~44, pointing to a colder water reservoir in PDS~70. This is consistent with the lower luminosity and lower accretion rate of PDS~70.}}
\label{fig4}
\end{figure}

\textit{Spitzer}-IRS observations detected water in $\sim$50\% of dust-rich inner disks around T~Tauri stars \cite{Pontoppidan2010}, but obtained only upper limits for disks with large inner dust gaps or cavities defined by a mid-IR spectral index $n_\mathrm{13-30}$ > 0.9, where $n$ is the slope of the spectrum between 13 and 30 $\mu$m (Fig.~\ref{fig4}).\cite{Banzatti2020}
The detection of water vapour in the PDS~70 MIRI spectrum demonstrates that PDS~70 has maintained to some degree the physical and chemical conditions of dust-rich inner disks in its terrestrial planet-forming zone despite the presence of a remarkably large gap (see `Origin of water in PDS~70' in Methods). Our LTE slab model only provides a first quantitative analysis of the H$_2$O emission. Non-LTE effects could lead to sub-thermal line emission which would make our estimated emitting area a lower limit. In T~Tauri disks with strong radial temperature gradients, the water lines are expected to originate from different regions of the disk depending on their upper energy level and Einstein-$A$ coefficients.\cite{Blevins2016} Detailed modelling using a realistic disk structure and including non-LTE effects such as IR radiative pumping is needed in the future to further constrain the distribution of H$_2$O across the inner disk. However, this first analysis already proofs that the inner disk of PDS~70 is rich in water and the inferred slab model parameters are roughly consistent with a detailed thermo-chemical model \cite{Portilla-Revelo2023} (Portilla-Revelo, priv. communication, 2023).

\change{The luminosity of the 17 $\mu$m water lines is two orders of magnitude weaker for PDS~70 than DoAr~44.\cite{Salyk2015,Banzatti2020}
DoAr~44 is a system with similar properties to PDS~70, but characterized by $n_\mathrm{13-30}$ $<0.9$ (Fig.~\ref{fig4}).} This result points to a colder water reservoir in PDS~70, and is consistent with the lower luminosity and lower accretion rate of PDS~70.\cite{Manara2014,Manara2019} This work opens a new window on the origin of water in protoplanetary disks by showing that MIRI-MRS can now detect very weak ($\lesssim5~$mJy) water lines in the innermost regions of disks with large gaps, and hence that the presence of water in the terrestrial planet-forming zone of dust-depleted inner disks is not as rare as previously thought.

\section*{Methods}
\label{methods}

\change{\noindent \textbf{PDS~70 system.} 
PDS~70 (V1032 Cen) is a K7-type star in the Upper Centaurus-Lupus subgroup (d=113.4$\pm$0.5~pc \cite{Gaiacollaboration2018}) in a late stage of accretion\cite{Skinner2022} with an estimated age of 5.4$\pm$1.0 Myr.\cite{Mueller2018} The disk around PDS~70\cite{Gregorio-Hetem2002,Metchev2004,Riaud2006} hosts two actively accreting protoplanets: PDS~70~b and PDS~70~c which reside in a $\sim54~$AU-annular gap between an inner and outer disk.\cite{Keppler2018,Haffert2019} The presence of an inner dusty disk in the PDS~70 system has been inferred from both near-IR scattered light and ALMA images.\cite{Long2018,Keppler2019,Benisty2021} The 855~$\mu$m dust continuum emission from the innermost disk regions is confined within the orbit of PDS~70~b ($\sim22~$AU; Extended Data Fig.~\ref{figA1}), putting an upper limit to the inner disk radial extent of $\sim$18~AU.\cite{Benisty2021} A population of small dust grains may be responsible for the observed inner disk emission although the current dust mass estimates could support the simultaneous presence of small and large dust grains.\cite{Dong2012}}

\smallskip
\noindent \textbf{Observations and data reduction.} The PDS~70 disk (CD-40-8434) was observed with MIRI\cite{miri_rieke2015PASP,Wright2015} on 1 August 2022 as part of the Guaranteed Time Observation (GTO) programme 1282 (PI: Th. Henning)  \change{with number 66}. 
The disk was observed in FASTR1 readout mode with a 4-point dither pattern in the negative direction for a total on-source exposure time of 4,132~s. 
The Medium Resolution Spectroscopy (MRS)\cite{Wells2015} mode was used, which has four Integral Field Units (IFUs). Each IFU (referred to as channel) covers a different wavelength range and splits the field of view into spatial slices. Calibration and processing of IFU observations produces 3-dimensional spectral cubes. The latter are used to extract a final spectrum covering the MIRI 4.9$-$22.5~$\mu$m range and is a composite of the four IFUs: channel 1 (4.9$-$7.65~$\mu$m; $R\sim$3400), channel 2 (7.51$-$11.71~$\mu$m; $R\sim$3000), channel 3 (11.55$-$18.02~$\mu$m; $R\sim$2400), and channel 4 (17.71$-$22.5~$\mu$m; $R\sim$1600). Each channel is in turn composed by three sub-bands: SHORT (A), MEDIUM (B), and LONG (C) leading to a total of twelve wavelength bands.

\smallskip \smallskip
\noindent We processed the PDS~70 data using a hybrid data reduction pipeline made from the combination of the \change{JWST Science Calibration pipeline\cite{Bushouse2022}} (v1.8.4) stages 1 to 3, with dedicated routines based on the Vortex Image Processing (VIP) package\cite{GomezGonzalez2017,Christiaens2023} 
for bad pixel correction, background subtraction and removal of spikes affecting the final spectrum. 
Specifically, data reduction proceeded as follow: (i) the class \texttt{Detector1} of the JWST pipeline was used to process uncalibrated raw data files using Calibration Reference Data System (CRDS) context \texttt{jwst\_1019.pmap} and default parameters; (ii) apart from pixels flagged in the Data Quality (DQ) extension, we identified additional bad pixels with both an iterative sigma clipping algorithm and through a cross-shaped match filter,  
and corrected them using a 2D Gaussian kernel; (iii) \texttt{Spec2} was then used with default parameters, but the background subtraction was skipped, and dedicated reference files\cite{Gasman2023} for photometric and fringe flat calibrations were adopted;
(iv) as no dedicated background observation was taken, we leveraged the four-point dither pattern to obtain a first guess on the background map, then refined it using a  
median-filter which both smoothed the background estimate and removed residual star signals from it; (v) \texttt{Spec3} was then run with default parameters, apart from the \texttt{master\textunderscore background} and \texttt{outlier\textunderscore detection} steps which were turned off, in the latter case to avoid spurious spectral features resulting from under-sampling of the Point Spread Function (PSF); (vi) we recentered the spectral cubes by applying the shifts maximizing the cross-correlation between cube frames, and found the location of the PSF centroid with a 2D Gaussian fit on the median image of \change{each} aligned cube; (vii) spectra were then extracted with aperture photometry in $2.5$-\change{FWHM} apertures centered on the centroid location (with the FWHM equal to $\sim$1.22 $\lambda/D$ with the telescope diameter $D$ equal to 6.5~m), corrected for both aperture size using correction factors\cite{Argyriou2023}, 
and spikes affecting individual spaxels included in the aperture; 
(viii) spectra were finally corrected for residual fringes at the spectrum level
and the bands were stitched together based on the level of the shorter wavelength bands (these rescaling factors were systematically within 3\% of the photometric solution). Spurious data reduction artefacts were masked at 
5.12$~\mu$m, 5.90$~\mu$m, 7.45$~\mu$m, 7.50$~\mu$m. The uncertainty associated to each photometric measurement considers both Poisson and background noise, combined in quadrature. The former is an output from the JWST pipeline, while for the latter we propagated our background estimate obtained in step (iv) through \texttt{Spec3}, and considered the standard deviation of the fluxes inferred in independent $2.5$-\change{FWHM} apertures as a proxy for the background noise uncertainty. The final relative uncertainties range from $\sim$0.1\% to $\sim$1.1\% with respect to the continuum at the shortest and longest wavelengths considered in this work (4.9 and 22.5$~\mu$m, respectively).

\smallskip
\noindent \textbf{Local continuum fit.} 
Extended Data Fig.~\ref{figA2} shows the local baseline fit for the 7~$\mu$m region. The continuum level is determined by selecting line-free regions and adopting a cubic spline interpolation (\texttt{scipy.interpolate.interp1d}). This continuum is then subtracted from the original data to produce the spectrum shown in Fig.~\ref{fig3}. 

\smallskip
\noindent \textbf{Correction for the photospheric emission.}
The observed near-IR colour index, $J - K_s = 1.01$ (2MASS), indicates a small colour excess, $E(J-K_s) \approx 0.16$ which could be due to either interstellar extinction or a true excess in the $K_s$ band, or a combination of the two. By assuming that the brightness in the $K_s$ band is essentially due to photospheric emission we can, by using a model atmosphere kindly provided by P. Hausschildt (personal communication, 2023), extrapolate the contribution into the mid-IR spectral region. The parameters used for the model atmosphere of the PDS~70 K7-star are an effective temperature $T_\mathrm{eff} = 4000~$K, surface gravity log$(g) = 4.5$, and solar metallicity.
At $5~\mu$m the photospheric contribution amounts to 56~mJy, i.e., 44\% of the observed flux density. 
At longer wavelengths, in the $7~\mu$m region where the water emission is detected, the photosphere amounts to one-third of the observed flux density and the subtraction of the photospheric contribution just marginally alters the continuum-subtracted spectrum (Extended Data Fig.~\ref{figA3}). 

\smallskip
\noindent \textbf{Slab models fits.} The molecular lines are analysed using a slab approach that takes
into account optical depth effects. The level populations are assumed
to be in Local Thermodynamical Equilibrium (LTE) and the line profile
function to be Gaussian with a full-width half maximum of
$\Delta V = 4.7$ km s$^{-1}$ ($\sigma = 2~$km/s).\cite{Salyk2011} The line emission is
assumed to originate from a slab of gas with a temperature $T$ and
a line of sight column density of $N$. Under these assumptions and
neglecting mutual line opacity overlap, the frequency-integrated intensity
of a line is computed as follows:

\begin{equation}
I=  \frac{\Delta V}{2 \sqrt{\ln 2} \lambda_0} B_{\nu_0}(T) \int_{-\infty}^{+\infty} \left (1-\exp (-\tau_0 e^{-y^2}) \right) dy,
\end{equation}
where $B_{\nu_0}(T)$ is the Planck function, $\lambda_0$ is the rest wavelength of the line, and $\tau_0$ is the optical depth at the line center $\nu_0$, with:
\begin{equation}
\tau_{0}  = \sqrt{\frac{\ln 2}{\pi}} \frac{A_{ul} N \lambda_0^3}{ 4 \pi  \Delta V} (x_{l} \frac{g_u}{g_l} - x_{u}).
\end{equation}
In this equation, $x_{l}$ and $x_{u}$ denote the level population of
the lower and upper states, and $g_l$ and $g_u$ \change{their respective
statistical weights, and $A_{ul}$ the spontaneous downward rate of the transition.} The line intensity is then converted into integrated flux $F_{\nu_0}$ assuming an effective emitting area of $\pi R^2$ and a distance to the source $d$ as:
\begin{equation}
F_{\nu_0}= \pi \left( \frac{R}{d} \right)^2 I.
\end{equation}
\change{We note that neglecting mutual line overlap for H$_2$O when calculating the line intensity is a valid approximation for N(H$_2$O)$\lesssim 10^{20}$cm$^{-2}$ and significantly reduces the computational time.\cite{Tabone2023}}
Finally, the spectrum is convolved and sampled in the same way as the observed spectrum\cite{Labiano2021}
and all lines are then summed to prepare a total synthetic spectrum. The molecular data, i.e., line positions, Einstein $A$
coefficients and statistical weights stem from a previous work.\cite{Tennyson2001}

\smallskip
\noindent \textbf{Fitting procedure for H$_2$O vapour lines.} The LTE slab model described above is then used to fit the H$_2$O lines in the 6.78$-$7.36$~\mu$m region following a $\chi^2$ method. First, an extended grid of models is computed varying the total column density from $10^{15}$ to $10^{20}$ cm$^{-2}$ in steps of 0.17 in log10-space and the temperature from 100$-$1400 K in steps of 50~K. We further assume an ortho-to-para ratio of 3. For each set of free parameters ($N, T, R$), a synthetic spectrum is calculated at the spectral resolving power $R = 2000$ and rebinned to the spectral sampling of the observed spectrum using the \texttt{slabspec} python code.\cite{Salyk2020} The adopted spectral resolution is lower that the nominal MIRI-MRS spectral resolution in channel 1;\cite{Wells2015} it was selected to account for the observed line broadening. This spectrum is further used to compute the $\chi^2$ value on a spectral channel basis. Specific spectral windows are chosen to avoid contamination by other gas features. This includes all spectral channels falling within 0.02~$\mu$m ($1000$~km/s) of any hydrogen recombination lines and a 0.01~$\mu$m wide spectral window at the position of the $S$(5) line of H$_2$ at 6.91~$\mu$m. In order to mitigate the errors induced by the continuum subtraction procedure, we also include only spectral elements falling within 0.004~$\mu$m of a water line. For each value of ($N$, $T$), the $\chi^2$ is then minimized by varying the emitting size $\pi R^2$. The resulting $\chi^2$ map is shown in Extended Data Fig.~\ref{figA4} together with the best fit emitting radius. The confidence intervals are estimated following a previous work\cite{Grant2023} and using a representative noise level of $\sigma=0.15~$mJy. We note that due to the large number of lines in this crowded region, there is little space to determine the noise on the continuum. Therefore, we estimate the noise level between $7.72~\mu$m and $7.73~\mu$m to avoid contamination by H$_2$O and hydrogen lines.

\smallskip
\noindent \textbf{Origin of water in PDS~70.} 
At the typical densities of inner disk regions ($n_\mathrm{H} \geq 10^8~$cm$^{-3}$) the chemistry can rapidly reach steady-state conditions and water vapour can form from a simple reaction sequence involving O, H$_2$, and OH. Water and OH absorb efficiently in the UV (i.e., water and OH shielding), ensuring the survival of water molecules even in regions of reduced dust opacity.\cite{Bethell2009} This mechanism by itself is able to account for the column densities of water vapour detected in this work and is supported by the presence of CO$_2$ emission. Small grains in the inner disk provide additional UV shielding. One question that naturally arises is whether the water vapour in PDS~70 originated prior to the formation of the giant protoplanets within the gap or whether there is a continuous supply of gas from the outer to the inner disk regions. ALMA high spatial resolution CO observations reveal the presence of gas inside the gap.\cite{Keppler2019,Facchini2021} Observations and models find the gap to be gas depleted\cite{Keppler2019} (2-3 orders of magnitude assuming an $r^{-1}$ surface density profile) and dust depleted\cite{Portilla-Revelo2022}, but not empty. One possibility could be that a population of water-containing dust particles is able to filter through the orbits of PDS~70~b and PDS~70~c, enriching the inner disk reservoir.\cite{Benisty2021} Experimental evidence suggests that water chemically bound to complex silicates can be preserved to temperatures up to $400-500~$K\cite{Zhuravlev2000,Stevenson2011} and thus survive in the regions probed by our observations inside the water snowline. We note that some degree of dust filtering is expected with gas replenishment, as small dust particles can couple to the gas. Therefore a replenishment of both gas and dust from the outer disk to sustain the water reservoir and hence PDS~70's accretion rate is possible.

\smallskip
\noindent \textbf{Fitting procedure for the dust continuum.}
The $4.9-22.5~\mu$m dust continuum is analysed using a two-layer disk model for the dust emission.\cite{Juhasz2009} This model was successfully applied to \textit{Spitzer}-IRS spectra of planet forming disks;\cite{Juhasz2010} we follow the same modeling approach here. We rebin the spectrum by averaging 15 spectral points and assign errors $\sigma$ to the rebinned spectral points assuming a normal error distribution with equal weights for each individual spectral element. The stellar photospheric emission is represented by a stellar atmosphere model fitted to optical and near-infrared photometry. The disk model has three spectral components: (1) a hot inner disk $F_{\rm rim}$, (2) an optically thick mid-plane disk layer $F_{\rm mp}$, and (3) an optically thin warm disk surface layer $F_{\rm sur}$. The dust grains representing the disk components are assumed to have power-law temperature distributions, and each is characterized by a minimum and a maximum temperature $T_\mathrm{atm}$. The disk surface layer is assumed to consist of a number of dust species $i$ with different chemical compositions, and with a fixed number of grain sizes $j$, all emitting at the same temperatures. The total disk flux can then be written as: 

\begin{equation}
    F_{\nu} = F_{\nu, \, \mathrm{rim}} + F_{\nu, \, \mathrm{mp}} + F_{\nu, \, \mathrm{sur}}
\end{equation}
where
\begin{equation}
    F_{\nu, \, \mathrm{sur}} = \sum_{i=1}^{n} \sum_{j=1}^{m} D_{i,j} \kappa_{i,j} \int_{T_\mathrm{atm,max}}^{T_\mathrm{atm,min}} \cfrac{2 \pi}{d^2} \, B_\nu (T) \, T^{\frac{2-q_\mathrm{atm}}{q_\mathrm{atm}}} dT
   \end{equation} 
and B$_{\nu}$($T$) is the Planck function, $q_\mathrm{atm}$ is the power law exponent for the temperature gradient in the disk surface layer, $\kappa_{i,j}$ are the opacities in cm$^2$g$^{-1}$ of dust species $i$ with grain size $j$, $d$ is the distance to the star, and D$_{i,j}$ are normalization factors.\cite{Juhasz2009} We use three grain compositions (with SiO$_2$, SiO$_3$, and SiO$_4$ stochiometry)\cite{Jager2003,Dorschner1995,Sogawa2006,Jager1998,Henning1997}  and both amorphous and crystalline lattice structures to capture the rich spectral structure evident in the MIRI data. The choice of this set of compositions is based on previous analyses,\cite{Juhasz2010} that showed that this set of materials is able to capture most spectral variations in planet-forming disks observed with \textit{Spitzer}-IRS. We use either 2 or 3 grain sizes (i.e., $0.1~\mu$m, $2~\mu$m, $5~\mu$m) for each of the dust species. In total the model has 23 fitting parameters. We use the MultiNest Bayesian fitting algorithm\cite{Feroz2008} and the PyMultiNest package\cite{Buchner2014} to find the best-fit parameters. The resulting fit and the separate spectral components (star, inner rim, midplane, surface layer) are shown in Fig.~\ref{fig2}. 

\smallskip
\noindent \textbf{WISE Time-series observations.}
Extended Data Fig.~\ref{figA6} reports WISE Time-series observations of PDS~70. Observations were executed on 2-3 February and 6 February 2010, and on August 1-2, 2010. We note that the source is highly variable and that WISE~4 (W4; $25\,\mu$m) is anticorrelated with WISE~1 (W1; $3.4\,\mu$m) and WISE~2 (W2; $4.6\,\mu$m). Such variability may be "seesaw"-like\cite{Muzzerolle2009}, where changes in the scale height of the inner disk wall shadow the disk material located further out. However, a complete "seesaw" profile is not observed, as at wavelengths shortwards of 8$~\mu$m the MIRI spectrum lies above the IRS spectrum (Fig.~\ref{fig1}). This is not surprising as the wavelength of the "pivot" point (i.e., the wavelength at which a shift in emission is observed) is dependent on the location of the occulting material with respect to the star, the stellar luminosity and the inclination of the system, with highly inclined systems showing a more complete "seesaw" than more face-on systems such as PDS~70 ($i=51.7 \pm 0.1^\circ$\cite{Keppler2019}). Interestingly, WISE~3 (W3; $12\,\mu$m) is not anticorrelated with WISE~1 and WISE~2 due to the dominant 10$~\mu$m silicate band which indeed shows a minor offset compared to the longer wavelengths. It indicates that the material contributing to the 10$~\mu$m emission is not shadowed. This behaviour is seen if the emission arises from warmer dust closer to the star than the occulting material, or further above the disk midplane. 

\smallskip
\noindent We also note that the difference in aperture size of \textit{Spitzer}-IRS and MIRI-MRS cannot explain the observed variability. The \textit{Spitzer}-IRS low-resolution spectrograph has a slit width of 3.6$''$ for wavelengths shortwards of 14$~\mu$m and 10.2$''$ for wavelengths longwards of 14~$\mu$m. While the maximum aperture of \textit{Spitzer}-IRS at longer wavelengths is larger than that of MIRI-MRS, this is  not the case for the shorter wavelengths where the slit widths are similar for both observatories ($\sim3.6''$ vs $4.0''$). However, a flux offset is also observed in this spectral region. Additionally, in the case that the long wavelength excess would arise from an extended component, a jump in flux level at 14~$\mu$m - where the aperture size changes - would be present in the IRS data, but it is absent.

\clearpage

\section*{Data Availability}
\change{The original data analysed in this work are part of the Guaranteed Time Observation (GTO) program  1282 (PI: Th. Henning) with number 66 and will become public on 2 August 2023 on the MAST database (\url{https://mast.stsci.edu}). The portion of the spectrum presented in Fig.~\ref{fig3} is available on Zenodo at \url{https://zenodo.org/record/7991022}. The spectroscopic data for water can be downloaded from the HITRAN database (\url{https://hitran.org}). The Spitzer-IRS spectrum plotted in Fig.~\ref{fig1} is part of the Spitzer-IRS GTO program 40679 (PI: G. Rieke). The spectrum was extracted and calibrated using private codes\cite{Bouwman2008,Juhasz2010} and is available on Zenodo at \url{https://zenodo.org/record/7991022}. The optical constants of the dust species considered in the fitting procedure for the dust continuum can be downloaded from the HJPDOC database (\url{https://www2.mpia-hd.mpg.de/HJPDOC}).} 

\section*{Code Availability}
\change{The slab model used in this work is a private code developed by B.T. and collaborators. It can be obtained from B.T. upon request. The synthetic spectra presented in this work can be reproduced using the slabspec code, which can be found at \url{https://doi.org/10.5281/zenodo.4037306}. The fitting procedure for the dust continuum uses the publicly available MultiNest Bayesian fitting algorithm (\url{https://github.com/JohannesBuchner/MultiNest}) and the PyMultiNest package (\url{https://github.com/JohannesBuchner/PyMultiNest}). Figures were made with Matplotlib version 3.5.1. under the Matplotlib license at \url{https://matplotlib.org/}.}

\section*{Acknowledgements}
The MINDS team would like to thank the entire MIRI European and US instrument team. Support from STScI is also appreciated. The following National and International Funding Agencies funded and supported the MIRI development: NASA; ESA; Belgian Science Policy Office (BELSPO); Centre Nationale d’Etudes Spatiales (CNES); Danish National Space Centre; Deutsches Zentrum fur L\"uft- und Raumfahrt (DLR); Enterprise Ireland; Ministerio De Economiá y Competividad; Netherlands Research School for Astronomy (NOVA); Netherlands Organisation for Scientific Research (NWO); Science and Technology Facilities Council; Swiss Space Office; Swedish National Space Agency; and UK Space Agency. G.P. would like to thank B. Bitsch and E. Gaidos for fruitful discussions and P. Hausschildt for kindly providing the model atmosphere. V.C. and O.A. acknowledge funding from the Belgian F.R.S.-FNRS. Th.H., R.F. and K.S. acknowledge support from the European Research Council under the Horizon 2020 Framework Program via the ERC Advanced Grant Origins 83 24 28. B.T. is a Laureate of the Paris Region fellowship program, which is supported by the Ile-de-France Region and has received funding under the Horizon 2020 innovation framework program and Marie Sklodowska-Curie grant agreement No. 945298. B.T. acknowledges support from the Programme National ‘Physique et Chimie du Milieu Interstellaire’ (PCMI) of CNRS/INSU with INC/INP cofunded by CNES. D.G. would like to thank the Research Foundation Flanders for co-financing the present research (grant number V435622N). D.G. and I.A. thank the European Space Agency (ESA) and the Belgian Federal Science Policy Office (BELSPO) for their support in the framework of the PRODEX Programme. I.K., A.M.A., and E.v.D. acknowledge support from grant TOP-1614.001.751 from the Dutch Research Council (NWO). I.K. and J.K. acknowledge funding from H2020-MSCA-ITN-2019, grant no. 860470 (CHAMELEON). E.F.v.D. acknowledges support from the ERC grant 101019751 MOLDISK and the Danish National Research Foundation through the Center of Excellence ``InterCat'' (DNRF150). T.P.R acknowledges support from ERC grant 743029 EASY. D.B. has been funded by Spanish MCIN/AEI/10.13039/501100011033 grants PID2019-107061GB-C61 and No. MDM-2017-0737. A.C.G. has been supported by PRIN-INAF MAIN-STREAM 2017 “Protoplanetary disks seen through the eyes of new generation instruments” and from PRIN-INAF 2019 “Spectroscopically tracing the disk dispersal evolution (STRADE)”. D.R.L. acknowledges support from Science Foundation Ireland (grant number 21/PATH-S/9339). L.C. acknowledges support by grant PIB2021-127718NB-I00,  from the Spanish Ministry of Science and Innovation/State Agency of Research MCIN/AEI/10.13039/501100011033.
\vspace{-10pt}  
\section*{Author contributions} 
G.P. and V.C. performed the data reduction, supported by D.G., M.S., I.A., and J.B. G.P., I.K., V.C., B.T., L.B.F.M.W., G.O. and Th.H. wrote the manuscript. G.P., B.T. and S.L.G. did the line analysis. L.B.F.M.W. carried out the dust continuum analysis. G.O. performed the correction for photospheric emission. J.B. performed the reduction of the \textit{Spitzer} dataset. Th.H. and I.K. planned and co-led the MINDS guaranteed time program. All authors participated in either the development and testing of the MIRI instrument and its data reduction, in the discussion of the results, and/or commented on the manuscript.
\vspace{-10pt}  
\section*{Competing Interests}
The authors declare no competing financial interests.
\vspace{-10pt}  
\section*{Correspondence} Correspondence and requests for materials should be addressed to G. Perotti.

\clearpage


\renewcommand\thefigure{\arabic{figure}}  
\section{Extended Data}\label{secA1}
\renewcommand{\figurename}{Extended Data Figure}
\setcounter{figure}{0}    

\begin{figure}[htb!]
\centering
\includegraphics[width=\hsize]{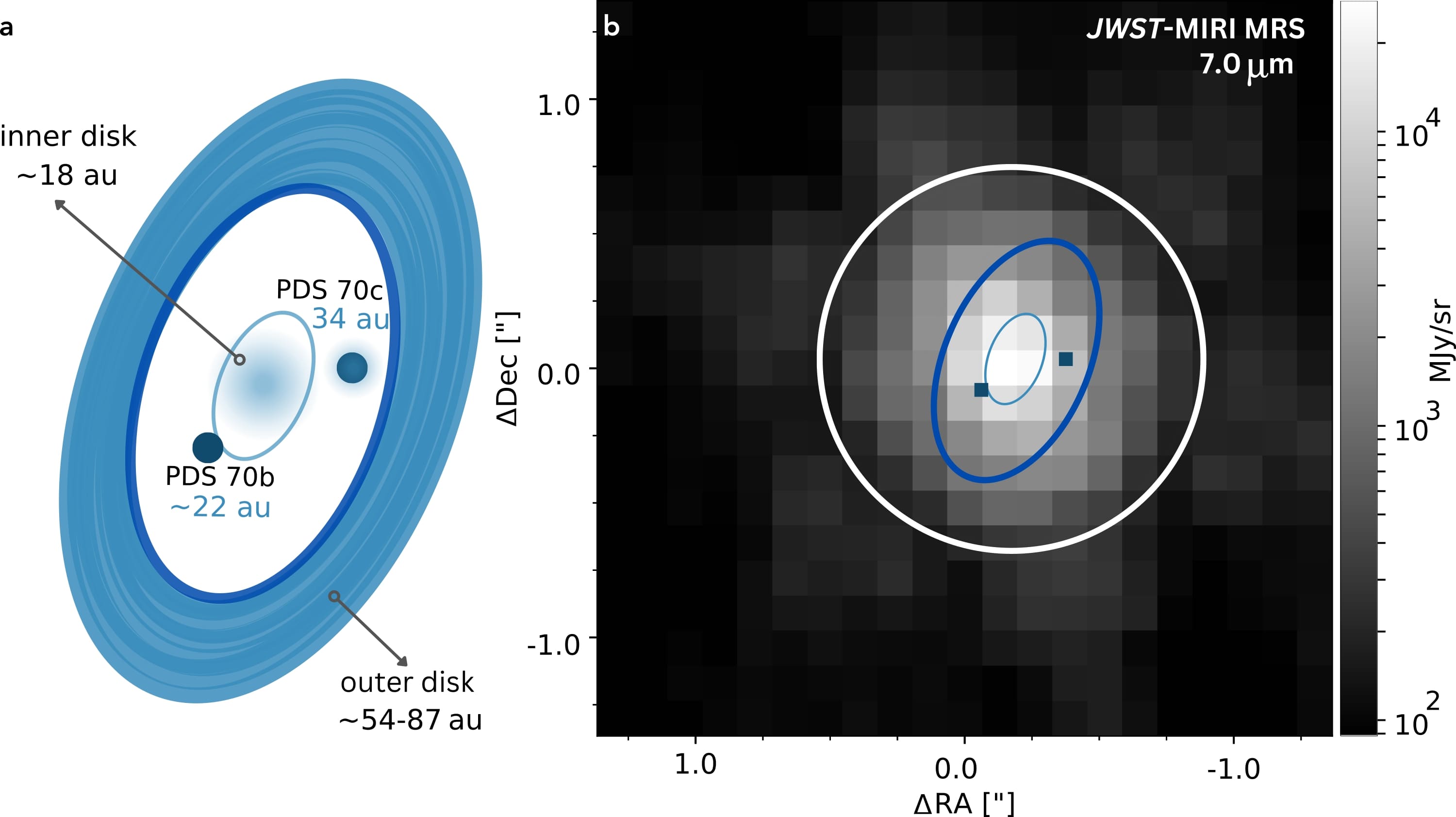}
 \caption{\textbf{The architecture of the PDS~70 system.} \textbf{a,} Schematic representation of the locations of the inner and outer disk of PDS~70 indicated as teal and blue ellipses. The protoplanets PDS~70~b and PDS~70~c are shown as blue dots. \textbf{b,} Main components of the schematic of the system on top of a MIRI-MRS IFU image at $7.0~\mu$m, illustrating the size of the system with respect to the 2.5-FWHM aperture used for spectro-photometric extraction (white circle). The latter linearly increases with wavelength.}
\label{figA1}
\end{figure}

\begin{figure}
\centering
\includegraphics[width=\hsize]{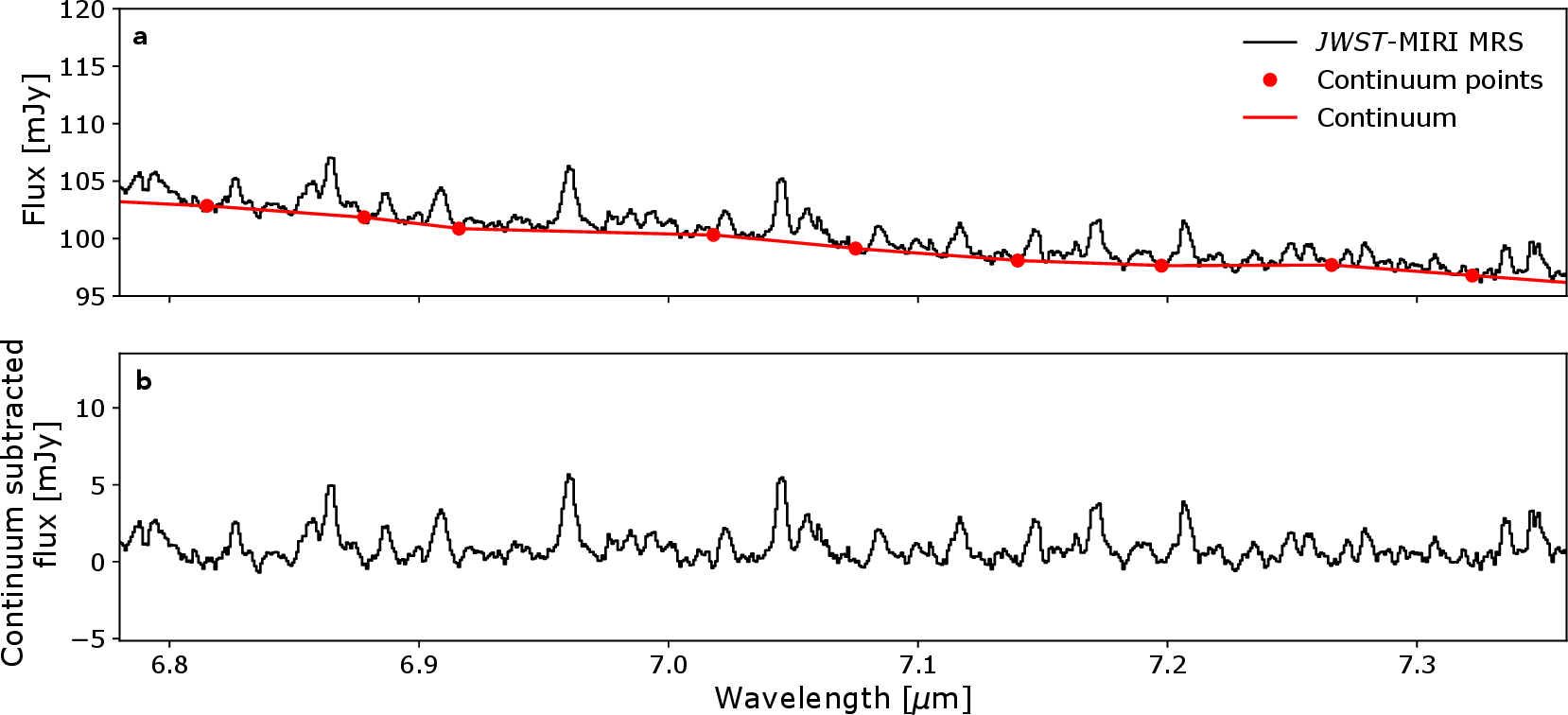}
 \caption{\textbf{Local continuum fit used in the spectrum presented in Figure~\ref{fig3}.} \textbf{a,} The selected continuum points are displayed as red dots and the interpolated continuum is shown as a red line. \textbf{b,} The continuum-subtracted spectrum.}
\label{figA2}
\end{figure}

\begin{figure}
\centering
\includegraphics[width=\hsize]{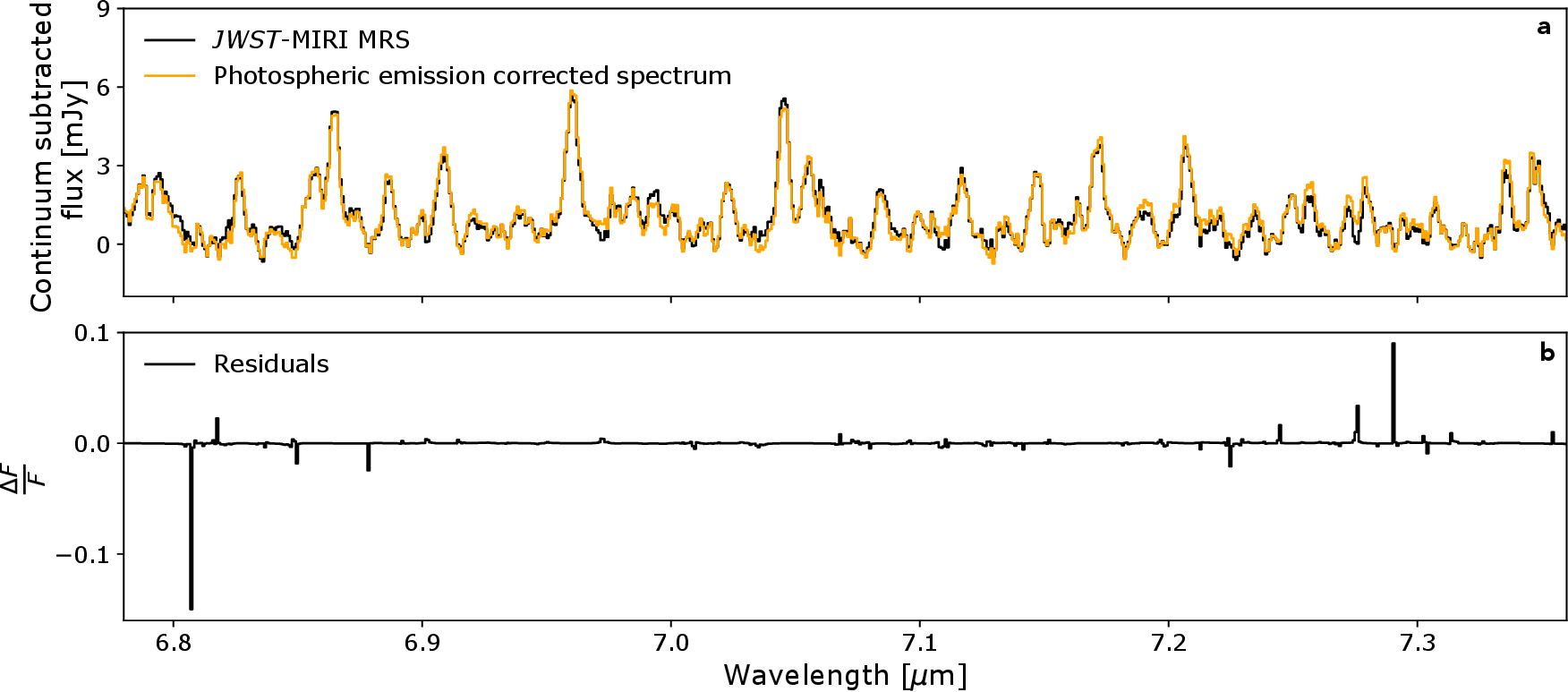}
\caption{\textbf{Correction for the photospheric emission. a} Comparison between the MIRI-MRS spectrum (black) and the spectrum corrected for the stellar photosphere (orange). Both spectra are continuum subtracted. \textbf{b,} \change{The residuals} show that the contamination from the stellar photosphere is negligible in the observed spectrum.}
\label{figA3}
\end{figure}

\begin{figure}
\centering
\includegraphics[width=\hsize]{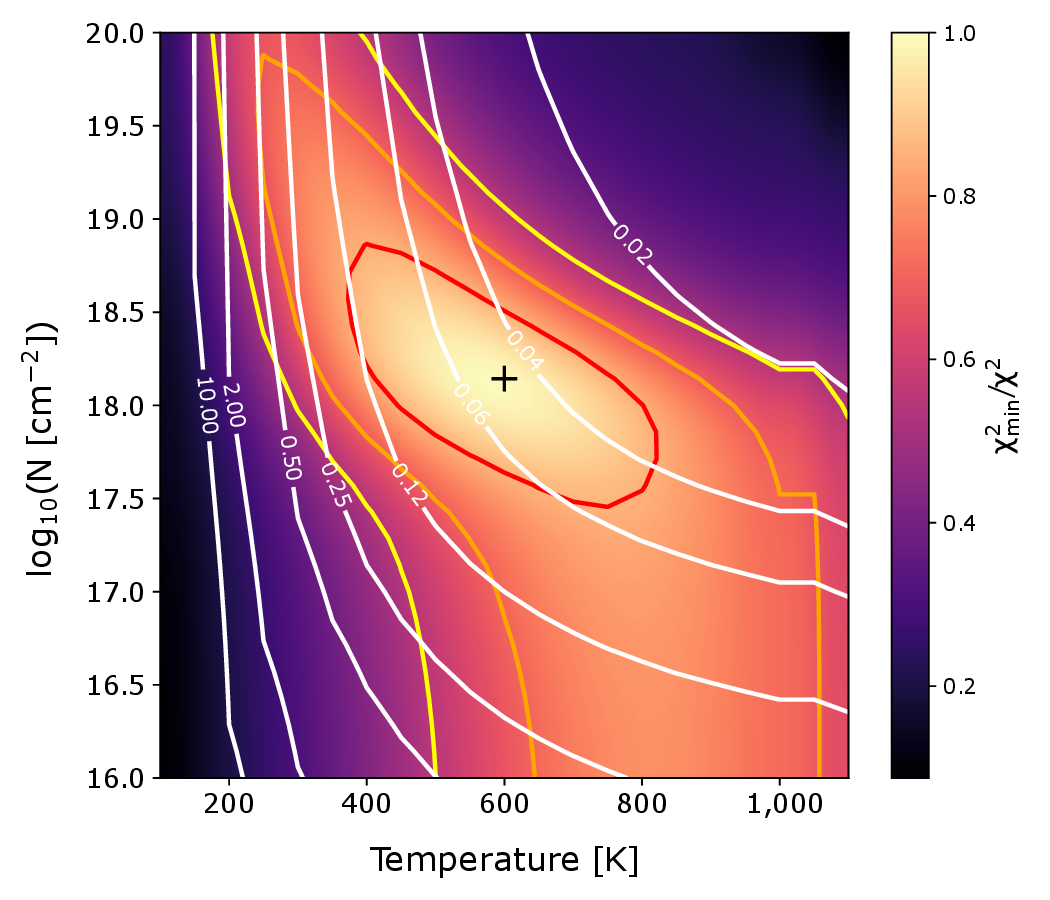}
\caption{\textbf{$\chi^2$ map for the fit of the 7$~\mu$m region of the H$_2$O bending mode.} The best-fit model is represented by a black plus. The $1 \sigma$, $2 \sigma$, and $3 \sigma$ confidence intervals are shown in red, orange, and yellow, respectively, for a typical noise level of $\sigma = 0.15$~mJy. The best-fitting emitting radius $R$ for all values of $N$ and $T$ is indicated as white lines. In general, we find a degeneracy between a high $T$ and low $N$ solution, and a low $T$ and high $N$ solution. Within the framework of our LTE slab model, the data indicate mildly optically thick H$_2$O emission at a temperature of about 600~K.}
\label{figA4}
\end{figure}

\begin{figure}
\centering
\includegraphics[width=\hsize]{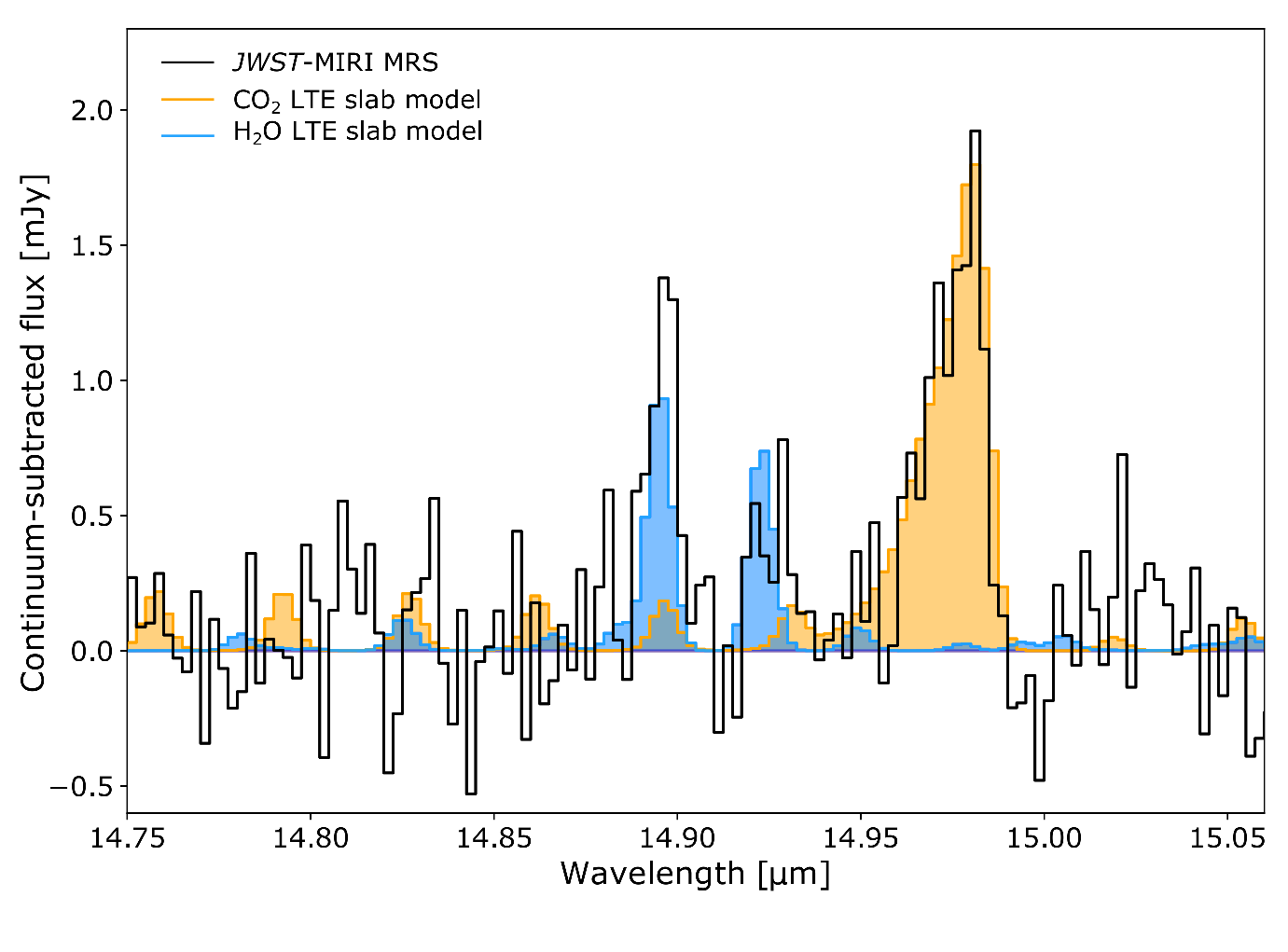}
\caption{\textbf{Continuum-subtracted spectrum in the $15~\mu$m region showing the detected Q-branch of CO$_2$ (orange).} The shape of this feature is sensitive to temperature and is well-fitted by an LTE slab model with $T \simeq 200$~K. The strength of this feature can be reproduced with $N \mathrm{(CO_2)} = 1.5\times10^{17}$~cm$^{-2}$ and $R=0.1~$AU. However, the aforementioned parameters are degenerate and are used for illustrative purposes only. Rotational lines of H$_2$O 
($J=14_{5\,\,\,\,\,\,10}-13_{2\,\,\,\,\,\,11},J=14_{6 \,\,\,\,\,\,9}-13_{3 \,\,\,\,\,\,10}$;
$E_u\sim4300~$K) 
are also detected and they are reasonably well reproduced by the best-fit model for the 7~$\mu$m region (blue). This could indicate that there is no additional reservoir of water at cooler temperature.}
\label{figA5}
\end{figure}

\begin{figure}
\centering
\includegraphics[width=\hsize]{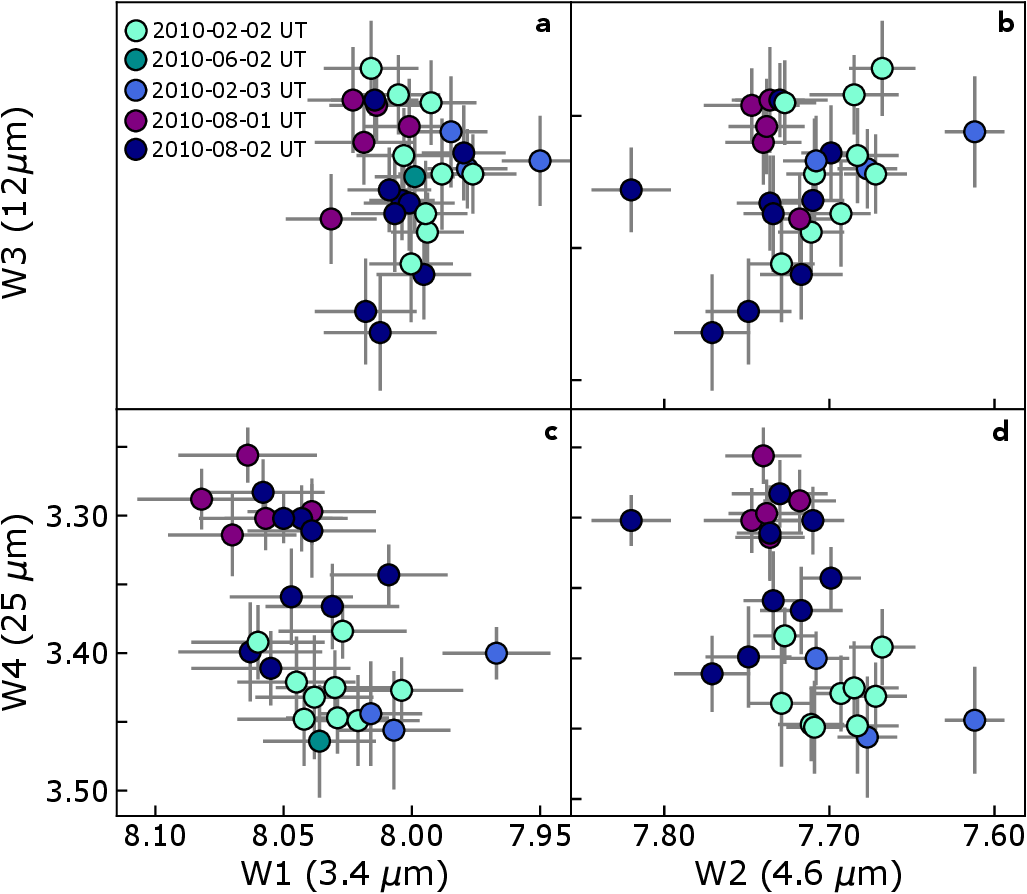}
\caption{\textbf{WISE Time-series photometry of PDS~70.} Errorbars represent 1 s.d. \textbf{a$-$b,} WISE~3 (W3) is not anticorrelated with WISE~1 and WISE~2 due to the dominant 10$~\mu$m silicate emission which does not vary substantially throughout different epochs. \textbf{c$-$d,} Anticorrelations are observed for WISE~1, WISE~2 and WISE~4 indicating a "seesaw"-like time variability \change{(see Methods for further details)}.}
\label{figA6}
\end{figure}

\clearpage


\end{document}